\documentclass[conference]{IEEEtran}
\usepackage{amsfonts}
\usepackage{graphicx}
\usepackage[cmex10]{amsmath}
\usepackage{algorithm}
\usepackage{algorithmic}
\usepackage{authblk}

\begin{document}

\title{Kronecker Product Correlation Model and Limited Feedback Codebook Design in a 3D Channel Model}

\author[1]{Dawei Ying}
\author[2]{Frederick W. Vook}
\author[2]{Timothy A. Thomas}
\author[1]{David J. Love}
\author[2]{Amitava Ghosh}
\affil[1]{School of Electrical and Computer Engineering, Purdue University, West Lafayette, IN}
\affil[2]{Nokia Solutions and Networks, Arlington Heights, IL}

\maketitle

\newtheorem{theorem}{Theorem}[section]
\newtheorem{proposition}[theorem]{Proposition}
\newtheorem{lemma}[theorem]{Lemma}
\newtheorem{remark}{Remark}

\def\argmin{\mathop{\mathrm{argmin}}}
\def\argmax{\mathop{\mathrm{argmax}}}
\def\diag{\mathop{\mathrm{diag}}}
\def\rank{\mathop{\mathrm{rank}}}
\def\opt{\mathop{\mathrm{opt}}}
\def\tr{\mathop{\mathrm{tr}}}
\def\vector{\mathop{\mathrm{vec}}}
\def\az{\mathop{\mathrm{az}}}
\def\el{\mathop{\mathrm{el}}}

\begin{abstract}
A 2D antenna array introduces a new level of control and additional degrees of freedom in multiple-input-multiple-output (MIMO) systems particularly for the so-called ``massive MIMO'' systems.  To accurately assess the performance gains of these large arrays, existing azimuth-only channel models have been extended to handle 3D channels by modeling both the elevation and azimuth dimensions. In this paper, we study the channel correlation matrix of a generic ray-based 3D channel model, and our analysis and simulation results demonstrate that the 3D correlation matrix can be well approximated by a Kronecker production of azimuth and elevation correlations. This finding lays the theoretical support for the usage of a product codebook for reduced complexity feedback from the receiver to the transmitter. We also present the design of a product codebook based on Grassmannian line packing.
\end{abstract}

\section{Introduction}
To meet the increasing demands on wireless communication systems, two-dimensional antenna arrays have been proposed for further improving the spectral efficiency of multi-input-multi-output (MIMO) technology.  A two dimensional antenna array provides control over not only the azimuth dimension, but the elevation dimension as well, thereby promising to further extend the gains from MIMO technology.  Various methods for controlling a 2D array have been proposed.  Elevation sectorization \cite{yilmaz2009system} and user-specific elevation beamforming \cite{thomas2012transparent} are two examples of how the additional diversity in a 3D channel can be exploited in current 4G LTE systems. In addition, Massive MIMO \cite{marzetta2010wireless} or Full-Dimension MIMO (FD-MIMO) \cite{nam2013full} operates with tens or even hundreds of antennas at the base station (BS) and enables the multiplexing of many users in a multi-user MIMO (MU-MIMO) fashion.

To accurately measure the performance of these 2D antenna arrays, a 3D channel model is needed where both the elevation and azimuth directions is are taken into account in the new model. At the time of writing this paper, 3GPP is actively developing a 3D channel model to enable the evaluation of elevation beamforming and massive MIMO. However, currently there are only extensions to the 2D 3GPP/ITU model, and where two examples are given in \cite{winner2010} and \cite{thomas20133d}. 

In a massive MIMO system, with a large number of antennas assembled within a limited space at the BS, the channels are highly likely to be correlated. Strong correlations will greatly reduce the effective degrees of freedom in wireless channels, which will significantly impact the performance of a massive MIMO system. In particular high correlation may make it difficult to send many spatial streams to one user, but may make simultaneous transmissions to a large number users (in a MU-MIMO fashion) from the 2D array practical. Therefore, the correlation statistics are useful for capacity analysis, and analytical expressions can provide insight to the correlation statistics of a full 3D channel. 

In this paper, we derive an analytic expression for correlation matrices with a generic ray-based non-line-of-sight (NLOS) 3D channel model. We compare the derived correlation matrix with the Kronecker product of correlations in azimuth and elevation dimensions. We found that even when a strictly mathematical equivalence does not hold, the eigenvalue distributions of two matrices, derived correlation and Kronecker product correlation, are surprisingly close to each other. Therefore, in an ergodic capacity analysis, the channel correlation matrices can be well approximated by the Kronecker product correlation model. This approximation indicates that it is possible to separate the 3D channel into azimuth and elevation directions and treat them as independent 2D channels for the purposes of designing an efficient feedback strategy and for designing MIMO transmit weights. 

Therefore, for a massive 2D antenna array with codebook-based feedback, instead of using a huge codebook for limited feedback, we can separately apply a product codebook, which is simply a Kronecker product of two smaller codebook designed for azimuth and elevation antenna dimensions. It is well-known that Grassmannian line packing is an important tool for optimal codebook design with both uncorrelated \cite{love2003grassmannian} and correlated channels \cite{love2006limited}. This paper examines the application of Grassmannian line packing to the design of a product codebook for operation in a 3D channel.  A specific product codebook design is presented.

\section{Channel Modeling}
\begin{figure}[!htb]
\includegraphics[width=\linewidth]{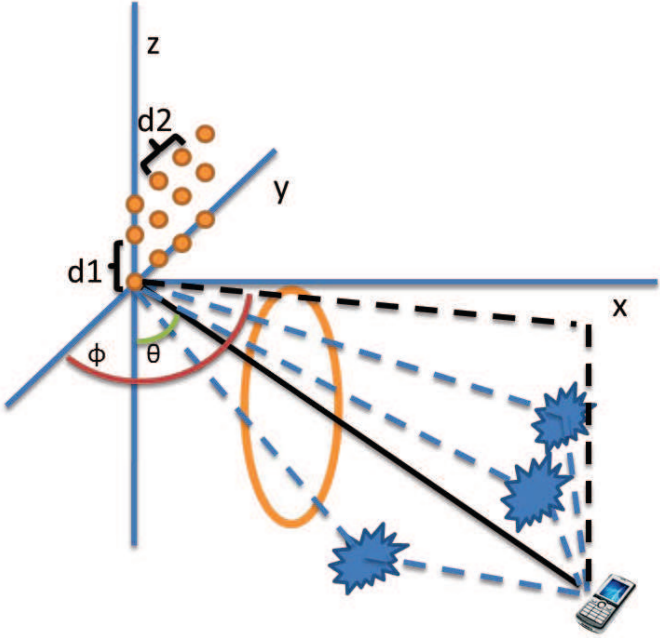}
\caption{Channel modeling with $L$ equal gain NLOS paths between the mobile and the base station, $M$ vertical antennas with $d_{1}$ wavelength spacing, and $N$ horizontal antennas with $d_{2}$ wavelength spacing. $\phi$ is for the azimuth angle, and $\theta$ is for the elevation angle}
\label{Fig:config}
\end{figure}

We consider a ray-based 3D channel model as shown in Figure \ref{Fig:config}. The mobile terminal is surrounded by local scatters, and the channel is assumed to consist of $L$ equal gain NLOS paths. Suppose the BS in y-z plane is equipped with a 2D antenna array with $M$ vertical antenna elements spaced by $d_{1}$ wavelengths, and $N$ horizontal antennas with a $d_{2}$ wavelength spacing. The BS array is deployed at a given height above the ground and typically the BS array will have some mechanical downtilt. For simplicity, we assume no mechanical downtilt for the antenna array in our model. The mobile terminal is assumed to have only one antenna for reception, but extensions to more than one mobile antenna are straightforward. Assuming a downlink transmission, let $\phi$ be the mean azimuth angle-of-departure (AoD), $\theta$ be the mean AoD in elevation, $\sigma$ be the standard deviation of azimuth angular perturbation, and $\xi$ be the standard deviation of elevation angular perturbation. For each path, we assume a random variable $\varphi$ to emulate the phase shift from the different lengths of the transmit paths. Note that a A similar 2D antenna array configuration was used in \cite{wang2012low} for angle-of-arrival (AoA) estimation.

The fast fading gain of path $k$ is represented by a random matrix for the 2D array given by
\begin{equation}
\mathbf{V}(\phi_{k}, \theta_{k}) = \left[
\begin{IEEEeqnarraybox*}[\relax][c]{c.c.c}
1 & \cdots & e^{-j(N-1)v_{k}}\\
e^{-ju_{k}} & \cdots & e^{-j[u_{k}+(N-1)v_{k}]}\\
\vdots & \ddots & \vdots\\
e^{-j(M-1)u_{k}} & \cdots & e^{-j[(M-1)u_{k}+(N-1)v_{k}]}
\end{IEEEeqnarraybox*}
\right]
\end{equation}
where 
\begin{equation}
u_{k} = \frac{2 \pi d_{1}}{\lambda} \cos\theta_{k} = \frac{2 \pi d_{1}}{\lambda} \cos(\theta + \Delta \theta_{k}),
\end{equation}
\begin{IEEEeqnarray*}{cl}
v_{k} &= \frac{2 \pi d_{2}}{\lambda} \sin \theta_{k} \cos\phi_{k} \\
&= \frac{2 \pi d_{2}}{\lambda} \sin (\theta + \Delta \theta_{k}) \cos(\phi + \Delta \phi_{k}). \IEEEyesnumber
\end{IEEEeqnarray*}
$\Delta \theta_{k}$ is the elevation angular perturbation for path $k$, and it is assumed to be normal distributed as $\mathcal{N}(0, \xi)$. Similarly, the azimuth angular perturbation $\Delta \phi_{k}$ is assumed to be distributed as $\mathcal{N}(0, \sigma)$. Moreover, assume angular perturbations $\Delta \phi_{i}$ and $\Delta \theta_{j}$ are independent variables for all $i = 1, \ldots, N$ and $j = 1, \ldots, M$. Note that elevation variance $\xi$ is a function of the distance between BS and mobile device, because recent research shows that the elevation spread has a strong distance dependence \cite{thomas20133d}.

Define
\begin{equation}
\mathbf{a}(u_{k}) = [1, e^{-j u_{k}}, \ldots , e^{-j (M-1) u_{k}}]^{T},
\end{equation}
\begin{equation}
\mathbf{b}(v_{k}) = [1, e^{-j v_{k}}, \ldots , e^{-j (N-1) v_{k}}]^{T}.
\end{equation}
We next rewrite $\mathbf{V}(\phi_{k}, \theta_{k}) = \mathbf{a}(u_k) \mathbf{b}^{T}(v_{k})$. With a random phase shift $\varphi$ uniformly distributed in $[0, 2 \pi]$, the channel response for the 2D array can be formulated as
\begin{equation}
\mathbf{H}(\phi, \theta, \sigma, \xi) = \sum\limits_{k = 1}^{L} \frac{e^{j \varphi_{k}}}{\sqrt{L}} \mathbf{a}(u_k) \mathbf{b}^{T}(v_{k}).
\end{equation}
Therefore, the channel vector is
\begin{equation}
\mathbf{h} = \vector(\mathbf{H}) = \sum\limits_{k = 1}^{L} \frac{e^{j \varphi_{k}}}{\sqrt{L}} \mathbf{b}(v_{k}) \otimes \mathbf{a}(u_{k}).
\end{equation}

In a full channel model, we need to specify more statistic parameters for the propagation. However, our ray-based channel model and the correlation we derive provide insight for the real 3D channel propagation. For example, the channel separability of a 3D channel into a Kronecker structure of an azimuth covariance matrix with an elevation correlation matrix..

\section{Channel Correlation: Analytical Expression}
In this section, we derive an analytical expression for the correlation matrix given the above 3D channel model. Although the derivation is given for this specific channel model, the methodology can be applied to any generic 3D channel model. Note that the random phase shift $\varphi_{k}$ is uniformly distributed, hence the mean of the channel vector is $0$.
\begin{equation}
\mathbb{E}\{\mathbf{h}(\phi, \theta, \sigma, \xi)\} = \sum\limits_{k = 1}^{L} \frac{\mathbb{E}\{e^{j \varphi_{k}}\}}{\sqrt{L}} \mathbb{E}\{\mathbf{b}(v_{k}) \otimes \mathbf{a}(u_{k})\} = 0.
\end{equation}

The correlation matrix, which is the same as the covariance matrix, can be calculated as
\begin{equation}
\mathbf{R}(\phi, \theta, \sigma, \xi) = \mathbb{E}\{\mathbf{h}(\phi, \theta, \sigma, \xi) \mathbf{h}^{H}(\phi, \theta, \sigma, \xi)\}.
\end{equation}
Since all propagation paths have equal gain and they are independent of each other, we need to consider only one arbitrary path and simplify the expression as
\begin{IEEEeqnarray*}{cl}
\mathbf{R} &= \sum\limits_{k = 1}^{L} \frac{1}{\sqrt{L}} \mathbb{E}\{(\mathbf{b}(v_{k}) \otimes \mathbf{a}(u_{k})) (\mathbf{b}(v_{k}) \otimes \mathbf{a}(u_{k}))^{H}\} \\
&= \mathbb{E}\{(\mathbf{b}(v) \otimes \mathbf{a}(u)) (\mathbf{b}(v) \otimes \mathbf{a}(u))^{H}\}.
\end{IEEEeqnarray*}
Next, we derive the expression for each entry of the correlation matrix. Define the $(k,l)$-th antenna element as the $k$-th in elevation and $l$-th in azimuth antenna in the 2D array, so it should be the $k+(l-1)M$-th element in the channel vector. Then the correlation between $(k,l)$-th and $(p,q)$-th antenna element is
\begin{equation}
\mathbf{R}_{(k, l), (p, q)} = \mathbb{E} \left\{ e^{ j \frac{2 \pi}{\lambda} A \cos(\theta + \Delta \theta + \eta)} \right\},
\end{equation}
where 
\begin{IEEEeqnarray*}{lcl}
A \cos(\theta + \Delta \theta + \eta) &=& (p-k) d_{1} \cos(\theta + \Delta \theta) \\
&+& (q-l) d_{2} \sin(\theta + \Delta \theta) \cos(\phi + \Delta \phi) \\
\IEEEyesnumber
\end{IEEEeqnarray*}
from the trigonometric identity. Let $\mu = \cos(\theta + \Delta \theta + \eta)$, and we approximate the distribution of $\mu$ as $\mathcal{N}(\bar{\mu}, \tilde{\xi})$ with 
\begin{equation}
\bar{\mu} = \cos(\theta + \eta)) \xi,
\end{equation}
and 
\begin{equation}
\tilde{\xi} = \sin(\theta + \eta) \xi.
\end{equation}
This type of approximation for the normal distribution using sine/cosine functions is commonly used in the propagation analysis of 2D channel modeling \cite{adachi1986cross}, \cite{zetterberg1995the}. Note that a different antenna ordering or an uplink transmission may lead to different signs for $(p-k)$ and $(q-l)$. However, it will not change the overall correlation matrix, since $\mathbf{R}$ is always conjugate-transpose symmetric.

Therefore, we will first take the expectation with respect to $\mu$, and integrate in the elevation direction to get
\begin{IEEEeqnarray*}{cl}
\mathbf{R}_{(k, l), (p, q)} &= \mathbb{E}_{\mu} \left\{ \frac{1}{\sqrt{2 \pi} \tilde{\xi}} \int_{-\infty}^{\infty} e^{j \frac{2 \pi }{\lambda} A \mu} e^ { -\frac{(\mu  - \bar{\mu})^{2}}{2 \tilde{\xi}^{2}} } d \mu \right\}\\
&= \mathbb{E}_{\mu} \left\{ e^{j \frac{2 \pi }{\lambda} A \bar{\mu}} e^{-\frac{1}{2} (\tilde{\xi} \frac{2 \pi }{\lambda} A)^{2}} \right\}. \IEEEyesnumber
\end{IEEEeqnarray*}
We can separate out the constant term, which happens to coincide with the correlation in elevation direction to get:
\begin{equation}
D_{1} = e^{j \frac{2 \pi d_{1}}{\lambda} (p-k) \cos\theta} e^{-\frac{1}{2} (\xi \frac{2 \pi d_{1}}{\lambda})^{2} (p-k)^{2} \sin^{2}\theta)}.
\end{equation}
To further simplify the expression, we define
\begin{equation}
D_{2} = \frac{2 \pi d_{2}}{\lambda} (q-l) \sin\theta,
\end{equation}
\begin{equation}
D_{3} = \xi \frac{2 \pi d_{2}}{\lambda} (q-l) \cos\theta,
\end{equation}
\begin{equation}
D_{4} = \frac{1}{2} \left( \xi \frac{2 \pi}{\lambda} \right)^{2} d_{1}d_{2}(p-k)(q-l) \sin(2 \theta).
\end{equation}
Hence, 
\begin{equation}
\mathbf{R}_{(k, l), (p, q)} = D_{1} \mathbb{E}_{\nu} \left\{ e^{j D_{2} \nu } e^{-\frac{1}{2} D_{3}^{2} \nu^{2} + D_{4} \nu } \right\},
\end{equation}
where $\nu = \cos(\phi + \Delta \phi)$ is approximately distributed as $\mathcal{N}(\cos\phi, \tilde{\sigma})$ with $\tilde{\sigma} = (\sin\phi) \sigma$. Then, we take the expectation with respect to $\nu$, and the correlation matrix is formulated as
\begin{IEEEeqnarray*}{cl}
\label{eqn:R}
\mathbf{R}_{(k, l), (p, q)} &= \frac{D_{1}}{\sqrt{2 \pi} \tilde{\sigma}} \int_{-\infty}^{\infty} e^{j D_{2} \nu } e^{-\frac{1}{2} D_{3}^{2} \nu^{2} + D_{4} \nu} e^{-\frac{(\nu - \cos(\phi))^{2}}{2 \tilde{\sigma}^{2}}} d \nu\\
&= \frac{D_{1}}{\sqrt{2 \pi} \tilde{\sigma}} \int_{-\infty}^{\infty} e^{j D_{2} \nu} e^{-\frac{ \left( \nu - \frac{D_{6}}{D_{5}} \right)^{2}}{\frac{2 \tilde{\sigma}^{2}}{D_{5}}} - \frac{D_{7}}{2D_{5}}} d \nu\\
&= \frac{D_{1}}{\sqrt{D_{5}}} e^{- \frac{D_{7}}{2D_{5}}} e^{ j \frac{D_{2} D_{6}}{D_{5}}} e^{ -\frac{1}{2} \frac{(D_{2} \tilde{\sigma})^{2}}{D_{5}}}, \IEEEyesnumber
\end{IEEEeqnarray*}
where 
\begin{eqnarray}
D_{5} = D_{3}^{2} \tilde{\sigma}^{2} + 1,
\end{eqnarray}
\begin{eqnarray}
D_{6} = D_{4} \tilde{\sigma}^{2} + \cos\phi,
\end{eqnarray}
\begin{eqnarray}
D_{7} = D_{3}^{2} \cos^{2}\phi - D_{4}^{2} \tilde{\sigma}^{2} - 2 D_{4} \cos\phi.
\end{eqnarray}

From the above analytical expressions, it is clear that term $D_{1}$ is only elevation related, i.e., only contains the term $(p-k)$, while $D_{2}, D_{3}$ and $D_{5}$ are azimuth related and only have the term $(q-l)$. Variable $D_{4}$, $D_{6}$, and $D_{7}$ have the cross term $(p-k)(q-l)$, contributing to both elevation and azimuth correlations. However, $D_{6}$ and $D_{7}$ are functions of $D_{4}$. Therefore if $D_{4} = 0$, the correlation term $\mathbf{R}_{(k, l), (p, q)}$ can be written as a product of elevation and azimuth correlations. Furthermore, if $D_{4}=0$ is true for all antenna index $k,l,p$ and $q$, then the correlation matrix is separable
\begin{equation}
\mathbf{R} = \mathbf{R}_{az} \otimes \mathbf{R}_{el},
\end{equation}
where
\begin{equation}
\label{eqn:R_el}
[\mathbf{R}_{el}]_{k,p} = e^{j \frac{2 \pi d_{1}}{\lambda} (p-k) \cos\theta} e^{-\frac{1}{2} (\xi \frac{2 \pi d_{1}}{\lambda})^{2} (p-k)^{2} \sin^{2}\theta)}
\end{equation}
denotes the elevation correlation, and the correlation in azimuth direction is
\begin{equation}
\label{eqn:R_az}
[\mathbf{R}_{az}]_{l,q} = \frac{1}{\sqrt{D_{5}}} e^{- \frac{D_{3}^{2} \cos^{2}\phi}{2D_{5}}} e^{ j \frac{D_{2} \cos\phi}{D_{5}}} e^{ -\frac{1}{2} \frac{(D_{2} \tilde{\sigma})^{2}}{D_{5}}}.
\end{equation}

\section{Kronecker Correlation Model}
The strictly mathematical separation discussed in the previous section is difficult to satisfy in general. Even if we relax the constraint to $D_{4} \approx 0$, an approximate separation is not likely to hold in many practical settings. For example, with a massive 2D antenna array, $D_{4} \approx 0$ for all possible $k,l,p$ and $q$ is a very harsh condition. For $D_{4} \approx 0$ to hold, we need either $\theta \approx \pi/2$ or $\sigma \approx 0$. $\theta \approx \pi/2$ means that the mobile device is very far away assuming there is no mechanical downtilt to the antenna array. In such a case, the elevation channel is less important, and the traditional 2D channel model suffices. Note also that $\sigma \approx 0$ indicates that the channel has a very small elevation angular spread. Given the distance dependence of the elevation spread, the small elevation spread may indicate that the device is very far away. Therefore, both cases do not provide much insight into the channel separable condition.

\subsection{Separation in Ergodic Capacity Analysis}
However, it is not necessary for us to have a strict requirement for a correlation model, such that every matrix entry is close to the analytical expression. For example, in a ergodic capacity analysis, if the eigenvalue distributions of two correlation matrices are close enough, then the resulting ergodic capacity will be similar. Therefore, we compare the eigenvalue distributions of the following correlation matrices: the analytical correlation matrix $\mathbf{R}$ defined as (\ref{eqn:R}), and the Kronecker correlation model
\begin{equation}
\mathbf{R}_{K} = \mathbf{R}_{\az} \otimes \mathbf{R}_{\el},
\end{equation}
with $\mathbf{R}_{\az}$ and $\mathbf{R}_{\el}$ defined in (\ref{eqn:R_az}) and (\ref{eqn:R_el}).

\begin{figure}[!htb]
\includegraphics[width=\linewidth]{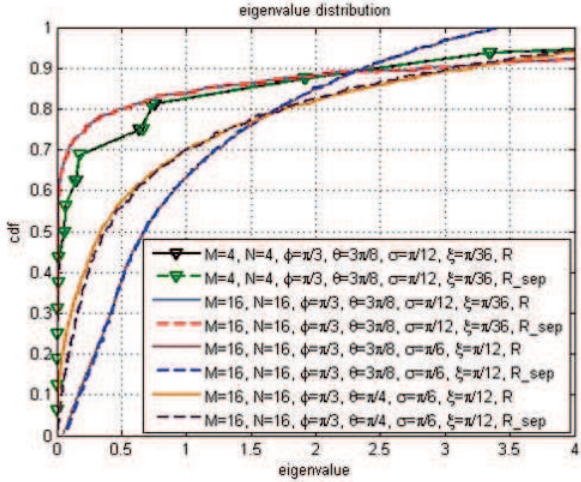}
\caption{The comparison on the eigenvalue distributions of two correlation matrices with various channel settings. Legends with ``R'' indicate the distribution for analytical correlation matrix $\mathbf{R}$, and ``R\_sep'' represents the distributions for Kronecker correlation model $\mathbf{R}_{K}$.}
\label{Fig:eigen_value}
\end{figure}

In Figure \ref{Fig:eigen_value}, we plot the eigenvalues distributions for 4 different channel configurations. We compare two correlation models for both moderate (4-by-4) 2D arrays and massive (16-by-16) 2D arrays. The comparison is obtained for both moderate ($\sigma = \pi / 12$ and $\xi = \pi / 36$) and large ($\sigma = \pi / 6$ and $\xi = \pi / 12$) angular spreads. The similarity in the eigenvalue distribution is then translated into the closeness of the capacity curves of using the both correlation matrices. 

\begin{figure}[!htb]
\includegraphics[width=\linewidth]{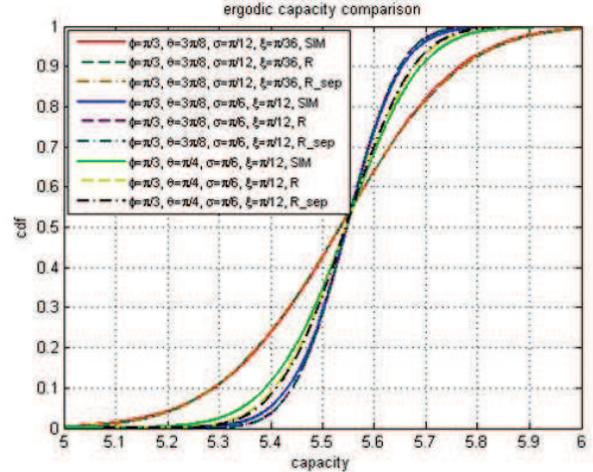}
\caption{The comparison on the ergodic capacity of channel realization and using two correlation models. Although there are some deviation in the capacity cdf curves between using correlation models and channel realization, two curves for correlation models are very close to each other in every simulation settings. Legends with ``SIM'' indicate the capacity curves are plotted using the channel model directly. Legends with ``R'' and ``R\_sep'' represent the curves using correlation matrices $\mathbf{R}$ and $\mathbf{R}_{K}$, respectively.}
\label{Fig:ergodic_capacity}
\end{figure}

In Figure \ref{Fig:ergodic_capacity}, we compare three ergodic capacity curves in each channel configuration. First, we generate the channel vector $\mathbf{h}$ following our ray-based NLOS 3D channel model, labled as ``SIM'' in the legend. We generate $L=20$ paths with randomly selected phase shift, and the azimuth and elevation AoDs selected from their respective distributions in the model. Second, we generate the channel vectors using the correlation matrices as
$\mathbf{h}_{\mathbf{R}} = \mathbf{R}^{1/2} \mathbf{w}$ and $\mathbf{h}_{\mathbf{R}_{K}} = \mathbf{R}^{1/2}_{K} \mathbf{w}$, where $\mathbf{w}$ is a random vector distributed as i.i.d $\mathcal{CN}(0,1)$. Let $\rho$ denote the signal-to-noise-ratio, and the ergodic capacity is then calculated as $C=\log(1+\rho \mathbf{h}^{H} \mathbf{h})$.
From the plots, we notice the correlation model curves have some visible deviation from the channel ``SIM'' results  because of the normal distribution approximation on the cosine functions. However, in all three different settings with a 16-by-16 2D antenna array, two correlation model curves are always on the top of each other.

The previous analysis indicates that massive 2D arrays with large elevation spread are likely to have a large $D_{4}$ term, and thus the correlation matrix is not mathematically separable. However, our simulation results demonstrate that even in cases with wide angular spread, which mean a large $D_{4}$ term, the Kronecker correlation model still has a similar eigenvalue distribution with as analytical expression. Therefore, the ergodic capacity performances of two correlation models only have little difference. 

\subsection{Separation in Feedback}
\begin{figure}[!htb]
\includegraphics[width=\linewidth]{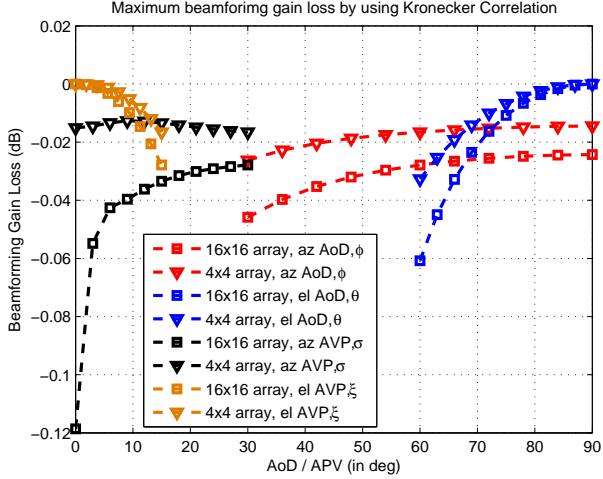}
\caption{The difference of maximum statistical beamforming gain by using full correlation matrix $\mathbf{R}$ and our Kronecker correlation model $\mathbf{R}_{K} = \mathbf{R}_{\az} \otimes \mathbf{R}_{\el}$. The gain loss is less than $0.06$ dB in most cases. The default setting of channel variables are : $\phi=\pi/3$, $\theta=3\pi/8$, $\sigma = \pi/6$ and $\xi=\pi/12$.}
\label{Fig:beamforming_gain}
\end{figure}

In this section, we show the channel separation property can be used in feedback. We approximate matrix $\mathbf{R}$ with
\begin{equation}
\mathbf{R} \approx \mathbf{R}_{\az} \otimes \mathbf{R}_{\el}.
\end{equation}
We first show the effect of using the Kronecker Correlation Model in statistical beamforming. Statistical beamforming \cite{raghavan2005when}, \cite{raghavan2007systematic} transmits signals along the dominant eigenvector of the correlation matrix in a spatial correlated channel. We assume $\lambda^{(1)}$ is the maximum eigenvalue for $\mathbf{R}$, and $\mathbf{u}^{(1)}$ is the corresponding dominant eigenvector. Similarly, we can define $\lambda^{(1)}_{\az}$, $\mathbf{u}^{(1)}_{\az}$, $\lambda^{(1)}_{\el}$ and $\mathbf{u}^{(1)}_{\el}$ for $\mathbf{R}_{\az}$ and $\mathbf{R}_{\el}$, respectively. From above ergodic capacity discussion, we know that the following approximation between eigenvalues holds
\begin{equation}
\lambda^{(1)} \approx \lambda^{(1)}_{\az} \lambda^{(1)}_{\el}.
\end{equation}
In fact, the approximation between corresponding eigenvectors also holds. If we use our Kronecker correlation model in statistical beamforming, the transmit signal is beamformed along $\mathbf{u}^{(1)}_{\az} \otimes \mathbf{u}^{(1)}_{\el}$ instead of $\mathbf{u}^{(1)}$. Hence, the maximum beamforming gain is
\begin{equation}
\mu = (\mathbf{u}^{(1)}_{\az} \otimes \mathbf{u}^{(1)}_{\el})^{H} \mathbf{R} (\mathbf{u}^{(1)}_{\az} \otimes \mathbf{u}^{(1)}_{\el}).
\end{equation}
In Figure \ref{Fig:beamforming_gain}, we plot the beamfoming gain loss by using $\mathbf{R}_{\az} \otimes \mathbf{R}_{\el}$. The simulations are conducted with various combination of channel variables: angle-of-departure (AoD) and angular perturbation variance (APV) for azimuth and elevation dimensions. The default setting for the variables are: $\phi=\pi/3$, $\theta=3\pi/8$, $\sigma = \pi/6$ and $\xi=\pi/12$, and we vary one variable in each simulation. From the figure, the loss is less than $0.12$ dB, and in most channel realizations, the loss is less than $0.06$ dB, which is negligible.

Figure \ref{Fig:beamforming_gain} demonstrates that $\mu$ is very close to $\lambda^{(1)}$, which means that $\mathbf{u}^{(1)}_{\az} \otimes \mathbf{u}^{(1)}_{\el}$, the dominant eigenvector of $\mathbf{R}_{K}$, is close to the dominant eigenvector of $\mathbf{R}$. Therefore, using our Kronecker model in statistical beamforming yields limited performance loss, and we separate the eigenvector feedback into azimuth and elevation directions. The separate feedback has smaller scale, i.e. a $\mathbb{C}^{N}$ vector and a $\mathbb{C}^{M}$ vector instead of a $\mathbb{C}^{MN}$ vector.

\begin{figure}[!htb]
\includegraphics[width=\linewidth]{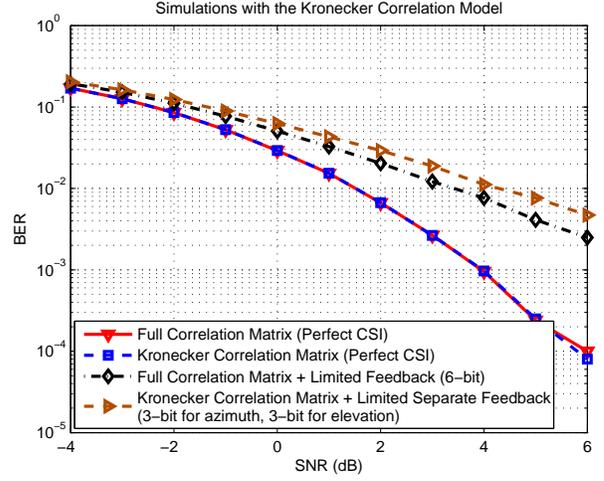}
\caption{Feedback performance for a 2-by-2 antenna array with $\phi=\pi/3$, $\theta=3\pi/8$, $\sigma = \pi/12$ and $\xi=\pi/36$. Two unlimited feedback schemes have almost identical performance. For limited feedback (Grassmannian line packing based), the separate feedback with the Kronecker correlation model suffers little performance loss compared, but it requires much smaller codebooks.}
\label{Fig:codebook_fb}
\end{figure}

Next, we show that the Kronecker correlation model is a good approximation with limited feedback. Consider the following beamforming transmission with input-output relationship
\begin{equation}
y = |\mathbf{h}^{H} \mathbf{f}| s + n,
\end{equation}
where $\mathbf{h} = (\mathbf{R}^{1/2})\mathbf{w}$ is the channel vector with $\mathbf{w}$ distributed as i.i.d $\mathcal{CN}(0,1)$, and $n$ is the $\mathcal{CN}(0,N_{0})$ noise. To maximize the receive SNR, the optimal infinite feedback beamformer is given by
\begin{equation}
\mathbf{f} = \frac{\mathbf{h}}{\|\mathbf{h}\|} = \frac{\mathbf{R}^{1/2}\mathbf{w}}{\|\mathbf{R}^{1/2}\mathbf{w}\|}.
\end{equation}
Suppose the correlation matrix is known to both end, and mobile user will feedback vector $\mathbf{w}$ to the BS. However, if we replace the $\mathbf{R}$ with our Kronecker model $\mathbf{R}_{K}$ at the BS, the beamformer is then formed as
\begin{equation}
\mathbf{f} = \frac{\mathbf{R}_{K}^{1/2}\mathbf{w}}{\|\mathbf{R}_{K}^{1/2}\mathbf{w}\|} = \frac{(\mathbf{R}_{\az} \otimes \mathbf{R}_{\el})^{1/2} \mathbf{w}}{\|(\mathbf{R}_{\az} \otimes \mathbf{R}_{\el})^{1/2} \mathbf{w}\|}.
\end{equation}
Figure \ref{Fig:codebook_fb} shows that these two unlimited feedback schemes have very close performance. For limited feedback, using our Kronecker correlation allow us to separate codebook and feedback with azimuth and elevation directions. In Figure \ref{Fig:codebook_fb}, the separate feedback for a 2-by-2 antenna array has a little performance degradation compare to conventional full feedback with same amount of feedback bits. However, with a 4-by-4 or an even larger 2D antenna array, constructing and maintaining a large codebook itself becomes a problem. Hence, separate feedback becomes a feasible solution for large antenna arrays. In next section, we present a product codebook design with our Kronecker correlation model.

\section{Product Codebook}
As shown in the previous section, assuming a separable channel correlation matrix leads to very small losses in beamforming gain. The Kronecker correlation model naturally leads to the idea of using a separate codebook for each of the azimuth and elevation dimensions. For massive MIMO, instead of applying a large $M \times N$ codebook according to the entire correlation matrix, we can construct two separate codebooks with elevation and azimuth correlation, respectively. Then we can take a Kronecker product of them to form the product codebook for MIMO system.
\begin{equation}
\mathcal{F}_{\az} \otimes \mathcal{F}_{\el} = \{c_{\az}^{(1)}, \ldots, c_{\az}^{(N_{1})}\} \otimes \{c_{\el}^{(1)}, \ldots, c_{\el}^{(N_{2})}\},
\end{equation}
the best entries are chosen as
\begin{IEEEeqnarray*}{cl}
(\mathbf{f}_{\az}, \mathbf{f}_{\el}) &= \argmax \limits_{\substack{\mathbf{f}_{\az} \in \mathcal{F}_{\az}\\\mathbf{f}_{\el} \in \mathcal{F}_{\el}}} \left|\mathbf{h}^{H} (\mathbf{f}_{\az} \otimes \mathbf{f}_{\el})\right|\\
&= \argmax \limits_{\substack{\mathbf{f}_{\az} \in \mathcal{F}_{\az}\\\mathbf{f}_{\el} \in \mathcal{F}_{\el}}} \left|\mathbf{w}^{H} (\mathbf{R}_{\az}^{1/2} \otimes \mathbf{R}_{\el}^{1/2}) (\mathbf{f}_{\az} \otimes \mathbf{f}_{\el})\right|\\
&= \argmax \limits_{\substack{\mathbf{f}_{\az} \in \mathcal{F}_{\az}\\\mathbf{f}_{\el} \in \mathcal{F}_{\el}}} \left|\mathbf{w}^{H} (\mathbf{R}^{1/2}_{\az} \mathbf{f}_{\az}) \otimes (\mathbf{R}^{1/2}_{\el} \mathbf{f}_{\el})\right|. \IEEEyesnumber
\end{IEEEeqnarray*}

\newcounter{tempequationcounter}
\begin{figure*}[!t]
\normalsize
\setcounter{tempequationcounter}{\value{equation}}
\begin{IEEEeqnarray*}{cll}
\label{eqn:grass_bound}
\setcounter{equation}{43}
d_{\mathrm{grass}}(\mathcal{F}) & \leq &2M_{t} \kappa^{1/2} \mathbb{E} [\min\limits_{i,k}\min\limits_{\phi, \theta} \|\mathbf{R}^{\frac{1}{2}} (e^{j\phi}\mathbf{w}_{\az} \otimes e^{j\theta}\mathbf{w}_{\el}) - \mathbf{R}^{\frac{1}{2}}  (\mathbf{c}_{\az}^{(i)} \otimes \mathbf{c}_{\el}^{(k)})\| ] \IEEEyessubnumber
\label{eqn:grass_bound_1}\\
& \leq &2M_{t} \kappa^{1/2} \mathbb{E} [\min\limits_{i,k}\min\limits_{\phi, \theta} \{\|\mathbf{R}^{\frac{1}{2}} _{\az} (e^{j\phi}\mathbf{w}_{\az}) \otimes \mathbf{R}^{\frac{1}{2}}_{\el}(e^{j\theta}\mathbf{w}_{\el} - \mathbf{c}_{\el}^{(k)}) \| + \|\mathbf{R}^{\frac{1}{2}}_{\az} (e^{j\phi}\mathbf{w}_{\az} - \mathbf{c}_{\az}^{(i)}) \otimes \mathbf{R}^{\frac{1}{2}}_{\el}(e^{j\theta} \mathbf{c}_{\el}^{(k)}) \|\}] \IEEEyessubnumber \label{eqn:grass_bound_2}\\
&=& 2M_{t} \kappa^{1/2} \mathbb{E} [\min\limits_{i,k}\min\limits_{\phi, \theta} \{\|\mathbf{R}^{\frac{1}{2}} _{\az} (e^{j\phi}\mathbf{w}_{\az})\| \|\mathbf{R}^{\frac{1}{2}} _{\el}(e^{j\theta}\mathbf{w}_{\el} - \mathbf{c}_{\el}^{(k)}) \| + \|\mathbf{R}^{\frac{1}{2}}_{\az} (e^{j\phi}\mathbf{w}_{\az} - \mathbf{c}_{\az}^{(i)})\| \|\mathbf{R}^{\frac{1}{2}} _{\el}(e^{j\theta} \mathbf{c}_{\el}^{(k)}) \|\} ]\\
&\leq& 2M_{t} \lambda_{\max} \lambda_{\min}^{-1/2} \mathbb{E} [\min\limits_{i,k}\min\limits_{\phi, \theta}  \{\|e^{j\phi}\mathbf{w}_{\az} - \mathbf{c}_{\az}^{(i)}\| + \|e^{j\theta}\mathbf{w}_{\el} - \mathbf{c}_{\el}^{(k)}\|\}] \IEEEyessubnumber \label{eqn:grass_bound_3}\\
&\leq& 2M_{t} \lambda_{\max} \lambda_{\min}^{-1/2} \{(2 - 2\mathbb{E}[\max\limits_{i} |\mathbf{w}_{\az}^{H} \mathbf{c}_{\az}^{(i)}|])^{1/2} + (2 - 2\mathbb{E}[\max\limits_{k} |\mathbf{w}_{\el}^{H} \mathbf{c}_{\el}^{(k)}|])^{1/2}\}.
\end{IEEEeqnarray*}
\setcounter{equation}{\value{tempequationcounter}}
\hrulefill
\vspace*{4pt}
\end{figure*}

To maximize the receive SNR, the optimal infinite Kronecker feedback vectors are given by
\begin{equation}
(\mathbf{w}_{\az}^{\opt}, \mathbf{w}_{\el}^{\opt}) = \argmax \limits_{\mathbf{w}_{\az}, \mathbf{w}_{\el}} \left|\mathbf{w}^{H} (\mathbf{R}^{1/2}_{\az} \mathbf{w}_{\az}) \otimes (\mathbf{R}^{1/2}_{\el} \mathbf{w}_{\el})\right|.
\end{equation}
Similarly to \cite{love2006limited}, we construct the azimuth codebooks
\begin{equation}
\mathcal{F}_{\az} = \left\{ \frac{\mathbf{R}^{1/2}_{\az}\mathbf{c}^{(1)}_{\az}} {\|\mathbf{R}^{1/2}_{\az}\mathbf{c}^{(1)}_{\az}\|} , \ldots, \frac{\mathbf{R}^{1/2}_{\az}\mathbf{c}^{(N_{1})}_{\az}} {\|\mathbf{R}^{1/2}_{\az}\mathbf{c}^{(N_{1})}_{\az}\|}\right\}
\end{equation}
and
\begin{equation}
\mathcal{F}_{\el} = \left\{ \frac{\mathbf{R}^{1/2}_{\el}\mathbf{c}^{(1)}_{\el}} {\|\mathbf{R}^{1/2}_{\el}\mathbf{c}^{(1)}_{\el}\|} , \ldots, \frac{\mathbf{R}^{1/2}_{\el}\mathbf{c}^{(N_{1})}_{\el}} {\|\mathbf{R}^{1/2}_{\el}\mathbf{c}^{(N_{1})}_{\el}\|}\right\}
\end{equation}
for elevation.

Therefore, to minimize the average SNR loss, we can formulate the distortion function as follows
\begin{IEEEeqnarray*}{cl}
&d_{\mathrm{grass}}(\mathcal{F}_{\az} \otimes \mathcal{F}_{\el}) \\
= &\mathbb{E}\left[\min\limits_{i,k} \left( \left|\mathbf{w}^{H} \frac{\mathbf{R}_{\az}\mathbf{w}_{\az}} {\|\mathbf{R}^{1/2}_{\az}\mathbf{w}_{\az}\|} \otimes \frac{\mathbf{R}_{\el}\mathbf{w}_{\el}} {\|\mathbf{R}^{1/2}_{\el}\mathbf{w}_{\el}\|} \right|^{2} \right.\right.\\
&-\left.\left. \left|\mathbf{w}^{H} \frac{\mathbf{R}_{\az}\mathbf{c}^{(1)}_{\az}} {\|\mathbf{R}^{1/2}_{\az}\mathbf{c}^{(1)}_{\az}\|} \otimes \frac{\mathbf{R}_{\el}\mathbf{c}^{(1)}_{\el}} {\|\mathbf{R}^{1/2}_{\el}\mathbf{c}^{(1)}_{\el}\|} \right|^{2} \right) \right]. \IEEEyesnumber
\end{IEEEeqnarray*}

We can bound the expression as in (\ref{eqn:grass_bound}), where $\lambda_{\max}$ denotes the largest eigenvalue of matrix $\mathbf{R}_{\az} \otimes \mathbf{R}_{\el}$, $\lambda_{\min}$ denotes its smallest eigenvalue, and $\kappa$ denotes its condition number. Inequality (\ref{eqn:grass_bound_1}) is given by \cite{love2006limited} and (\ref{eqn:grass_bound_2}) use the triangle inequality. Inequality (\ref{eqn:grass_bound_3}) comes from the fact that for all unit vectors $\mathbf{u}$ and $\mathbf{v}$
\begin{equation}
\left\|\mathbf{R}^{\frac{1}{2}}_{\az}\mathbf{u}\right\|\left\|\mathbf{R}^{\frac{1}{2}}_{\el}\mathbf{v}\right\| \leq \lambda^{\frac{1}{2}} _{\max} \|\mathbf{u}\|\|\mathbf{v}\| = \lambda^{\frac{1}{2}} _{\max}
\end{equation}

As shown in \cite{love2003grassmannian}, the upper bound in (\ref{eqn:grass_bound}) can be minimized by using Grassmannian line packing to generate two sub-codebooks $\left\{\mathbf{c}^{(1)}_{\az}, \ldots, \mathbf{c}^{(N_{1})}_{\az}\right\}$ and $\left\{\mathbf{c}^{(1)}_{\el}, \ldots, \mathbf{c}^{(N_{2})}_{\el}\right\}$. Figure \ref{Fig:codebook_fb} has a simulation results for the Grassmannian line packing based product codebook for a 2-by-2 antenna array. 

\section{Conclusion}
In this paper, we derived an analytic expression of the correlation matrix for a 2D antenna array using a ray-based 3D channel model. We demonstrated that the Kronecker correlation model has very similar eigenvalue distribution as the correlation matrix, and thus it is a good approximation for the original correlation matrix. Therefore the 3D channel can be separated into azimuth and elevation directions. Based on the channel separability, we presented a product codebook design using Grassmannian line packing.

\bibliographystyle{IEEEtran}
\bibliography{IEEEabrv,reference}{}

\end{document}